\documentclass[journal]{IEEEtran}
\makeatletter\if@twocolumn\PassOptionsToPackage{switch}{lineno}\else\fi\makeatother

\usepackage{graphicx}
\usepackage[T1]{fontenc}
\ifCLASSINFOpdf
\else
\fi
\usepackage{amsmath}
\usepackage[cmintegrals]{newtxmath}
\usepackage{url,multirow,morefloats,floatflt,cancel,tfrupee}
\makeatletter

\AtBeginDocument{\@ifpackageloaded{textcomp}{}{\usepackage{textcomp}}}
\makeatother
\usepackage{colortbl}
\usepackage{xcolor}
\usepackage{pifont}
\usepackage[nointegrals]{wasysym}
\makeatletter

\def\mcWidth#1{\csname TY@F#1\endcsname+\tabcolsep}

\def\cAlignHack{\rightskip\@flushglue\leftskip\@flushglue\parindent\z@\parfillskip\z@skip}
\def\rAlignHack{\rightskip\z@skip\leftskip\@flushglue \parindent\z@\parfillskip\z@skip}

\usepackage{ifxetex}
\ifxetex\else\if@twocolumn\usepackage{dblfloatfix}\fi\fi

\AtBeginDocument{
	\expandafter\ifx\csname eqalign\endcsname\relax
	\def\eqalign#1{\null\vcenter{\def\\{\cr}\openup\jot\m@th
			\ialign{\strut$\displaystyle{##}$\hfil&$\displaystyle{{}##}$\hfil
				\crcr#1\crcr}}\,}
	\fi
}

\AtBeginDocument{%
	\@ifpackageloaded{endfloat}%
	{\renewcommand\efloat@iwrite[1]{\immediate\expandafter\protected@write\csname efloat@post#1\endcsname{}}}{\newif\ifefloat@tables}%
}%

\def\BreakURLText#1{\@tfor\brk@tempa:=#1\do{\brk@tempa\hskip0pt}}
\let\lt=<
\let\gt=>
\def\processVert{\ifmmode|\else\textbar\fi}

\@ifundefined{subparagraph}{
	\def\subparagraph{\@startsection{paragraph}{5}{2\parindent}{0ex plus 0.1ex minus 0.1ex}%
		{0ex}{\normalfont\small\itshape}}%
}{}

\newcommand\role[1]{\unskip}
\newcommand\aucollab[1]{\unskip}

\@ifundefined{tsGraphicsScaleX}{\gdef\tsGraphicsScaleX{1}}{}
\@ifundefined{tsGraphicsScaleY}{\gdef\tsGraphicsScaleY{.9}}{}
\def\checkGraphicsWidth{\ifdim\Gin@nat@width>\linewidth
	\tsGraphicsScaleX\linewidth\else\Gin@nat@width\fi}

\def\checkGraphicsHeight{\ifdim\Gin@nat@height>.9\textheight
	\tsGraphicsScaleY\textheight\else\Gin@nat@height\fi}

\def\fixFloatSize#1{}
\let\ts@includegraphics\includegraphics

\def\inlinegraphic[#1]#2{{\edef\@tempa{#1}\edef\baseline@shift{\ifx\@tempa\@empty0\else#1\fi}\edef\tempZ{\the\numexpr(\numexpr(\baseline@shift*\f@size/100))}\protect\raisebox{\tempZ pt}{\ts@includegraphics{#2}}}}

\AtBeginDocument{\def\includegraphics{\@ifnextchar[{\ts@includegraphics}{\ts@includegraphics[width=\checkGraphicsWidth,height=\checkGraphicsHeight,keepaspectratio]}}}

\DeclareMathAlphabet{\mathpzc}{OT1}{pzc}{m}{it}

\def\URL#1#2{\@ifundefined{href}{#2}{\href{#1}{#2}}}

\def\UrlOrds{\do\*\do\-\do\~\do\'\do\"\do\-}%
\g@addto@macro{\UrlBreaks}{\UrlOrds}

\edef\fntEncoding{\f@encoding}

\makeatother

\newif\ifmultipleabstract\multipleabstractfalse%
\usepackage{tabulary}
\makeatletter
\AtBeginDocument{\@ifpackageloaded{longtable}{%
		\def\LT@makecaption#1#2#3{%
			\LT@mcol\LT@cols c{\hbox to\z@{\hss\parbox[t]\LTcapwidth{%
						\sbox\@tempboxa{#1{#2: } #3}%
						\ifdim\wd\@tempboxa>\hsize
						#1{#2: }\textsc{#3}%
						\else
						\hbox to\hsize{\hfil\box\@tempboxa\hfil}%
						\fi
						\endgraf\vskip\baselineskip}%
					\hss}}}
	}{}}
\makeatother

\makeatletter
\def\fig@textbf{\textbf}
\AtBeginDocument{\renewcommand\floatc@plain[2]{\setbox\@tempboxa\hbox{{\footnotesize#1.}\footnotesize\hskip.5em#2}%
		\ifdim\wd\@tempboxa>\hsize {\fig@textbf{\footnotesize#1.}}\footnotesize\hskip.5em#2\par
		\else\hbox to\hsize{\hfil\box\@tempboxa\hfil}\fi}}
\makeatother

\usepackage[flushleft]{threeparttablex}

\usepackage{float}

\newcommand{\texttildeapprox}{{\fontfamily{pcr}\selectfont\texttildelow}}

\begin{document}

%

        \title{\textbf{On the Convergence of Blockchain and Internet of Things (IoT) Technologies}}
      
\author{Mohammad~Maroufi, Reza~Abdolee, and	Behzad~mozaffari tazekand
	\thanks{Mohammad~Maroufi, Behzad~mozaffari tazekand are with WiLab, Faculty of Electrical and Computer Engineering, University of Tabriz, 29 Bahman Blvd., University of Tabriz, Tabriz, East Azarbayjan, Iran, e-mail: m.maroufi@tabrizu.ac.ir (Corresponding author: Mohammad~Maroufi) \& mozaffary@tabrizu.ac.ir}\thanks{Reza~Abdolee is with Wireless Communication Lab, Bakersfield, California State university , Science building III, Bakersfield, CA, USA, e-mail: rabdolee@csub.edu}}

\maketitle 

\begin{abstract}
The Internet of Things (IoT) technology will soon become an integral part of our daily lives to facilitate the control and monitoring of processes and objects and revolutionize the ways that human interacts with the physical world. For all features of IoT to become fully functional in practice, there are several obstacles on the way to be surmounted and critical challenges to be addressed. These include, but are not limited to cybersecurity, data privacy, energy consumption, and scalability. The Blockchain decentralized nature and its multi-faceted procedures offer a useful mechanism to tackle several of these IoT challenges. However, applying the Blockchain protocols to IoT without considering their tremendous computational loads, delays, and bandwidth overhead can let to a new set of problems. This review evaluates some of the main challenges we face in the integration of Blockchain and IoT technologies and provides insights and high-level solutions that can potentially handle the shortcomings and constraints of both IoT and Blockchain technologies.
\end{abstract}
    
%
\IEEEpeerreviewmaketitle

\section{Introduction}
Internet of Things (IoT) enables a network of physical objects (things), empowered by sensing, processing and communication units, to sense physical events, exchange data and interact with the environment to accordingly make decisions or monitor some processes and events without human interventions. One of the prominent motivation behind the advent of  IoT systems was to facilitate the real-time data collection and to provide automatic and remote control mechanisms replacing the today's conventional monitoring and control systems across different industries, such manufacturing, environmental monitoring, digital agriculture, smart cities and home, business management and asset tracking \cite{rayes2016internet}. It is predicted that by 2020, the number of connected devices surpasses 20 billion \cite{hung2017leading}. This growing demands and the tremendous expansion of IoT across emerging industries requires swift advancement in the current IoT protocols, technologies, and architectures and substantial progress in identifying the supporting IoT standards.

IoT systems generate massive volumes of data that require network connectivity and power, processing and storage resources to transform these data into meaningful information or services. Beside reliable connectivity and network scalability, cybersecurity and data privacy of are crucial importance in using IoT networks. Currently, centralized architecture models widely used to authenticate, authorize and connect different nodes in an IoT network. With the growing number of devices to hundreds of billions, centralized systems will break down and fail when the centralized server becomes unavailable. Decentralized IoT architecture was proposed to solve this issue, in which it moves away some of the network processing tasks to the edge \cite{ai2018edge}. For instance, in fog computing models, some of the critical operations that used to be processed by cloud servers are now assigned to be performed by IoT hubs or fog \cite{alrawais2017fog}. Peer-to-peer (P2P) architecture provides another solution, where neighboring devices directly interact with each other in meshes to identify, authenticate and exchange information without using any centralized node or agent between them \cite{buyya2016internet}.

IoT devices include both resource-constrained and resource-rich devices. Although some IoT devices such as smartphones and Raspberry Pi utilize sufficient resources, most of them feature limited power, processing, and memory resources due to their small sizes and low inherent design cost. Therefore, IoT devices and their protocols have to be designed to be resource efficient and meanwhile perform real-time processing, keep connectivity and protect the security and privacy of the transmit data \cite{haroon2016constraints,musaddiq2018survey}.

The battery capacity and computing power limitation created an obstacle to executing heavy and advanced cryptography algorithms to protect information. Critical security and privacy issues may arise in IoT devices because of sensitive personal data which connected things/objects reveal about their owners behavior and activities. Collecting such crucial data in centralized untrusted entities may create a significant privacy risk. This is probable in practice. For instance, Edward Snowden revealed that the PRISM program which operates under the United States National Security Agency (NSA), collects the data generated from Internet communications from various U.S. Internet providers \cite{conoscenti2016blockchain,yang2017survey}.

Due to the critical role of IoT devices in sensing the surrounding world and activating appropriately, collecting reliable data has a vital bearing on the precise functionality of these devices. IoT data reliability can be achieved by using distributed signal processing methods which execute a verification process among all its participants to ensure that data remain immutable and untampered. Considering this and understanding the basic features of Blockchain technology, which used as a cornerstone of Bitcoin \cite{nakamoto2008bitcoin}, we can intuitively find out the potential that Blockchain can offer to address the data reliability challenge in IoT. Bitcoin is supported by the Blockchain protocol to ensure that the information remains immutable. This protocol was proposed by a group of researchers in 1991 to timestamp digital documents and makes it impossible to backdate or tamper with them \cite{bashir2017mastering}.

The Blockchain suggests a way to record transactions or any digital interaction that is designed to be secure, transparent, highly resistant to outages, auditable, and efficient. These features encourage IoT companies to enhance their IoT network-based to Blockchain-based technology. It is a distributed ledger which managed by a peer-to-peer network to provides inter-node communication and verifying new blocks. Security may be considered as one of the most valuable features of the Blockchain. Once data recorded in the Blockchain, it cannot be modified without modification of all subsequent blocks and that needs a consensus of the network majority. The consensus algorithms used in Blockchain slow down the creation of new blocks and make it hard to tamper with previous blocks \cite{mougayar2016business}.
An intelligence convergence of IoT and Blockchain technologies can lead to a verifiable, secure and robust mechanism of storing and managing data generated or processed by smart connected devices. This network of interconnected devices will be able to interact with their environment and make decisions without any human intervention \cite{wood2018blockchain}.

Although, integrating Blockchain technology in IoT will enhance security, data privacy, and reliability of IoT devices, it create a new set of challenges. In recent years, researchers have widely studied the integration approaches, benefits, and challenges in IoT devices and networks \cite{conoscenti2016blockchain,reyna2018Blockchain,atlam2018Blockchain}. A  detailed literature review on the Blockchain applications in IoT was reported in \cite{conoscenti2016blockchain}, where the authors categorize previous research based on the Blockchain use cases in IoT and discuss their advantage and disadvantages. In \cite{reyna2018Blockchain,atlam2018Blockchain} researchers have introduced the key features of Blockchain, their application in IoT devices, challenges, and solutions to overcome challenges in IoT technology. Some investigations focused on the Blockchain distinct features and their impacts in integration with IoT. \cite{he2018consensus} reviewed the main consensus mechanisms and pointed out their strengths and weaknesses in IoT applications. They provided a comprehensive guide for developers to choose and design consensus mechanisms for Blockchain based consensus algorithms for IoT applications by considering their limited resources. In \cite{dorri2017Blockchain} investigators have analyzed the security and data privacy issues of IoT networks and explored Blockchain protocols as one of the potential solutions. Smart contracts and their roles in efficient controlling of IoT devices and related cybersecurity issues were discussed in \cite{christidis2016Blockchains}.

The aim of this paper is to present a comprehensive study on the integration of Blockchain and IoT and analyze different aspects of these embedded technologies. We attempt to provide strategic and technical insights into IoT restrictions and challenges, Blockchain specification and weaknesses, Blockchain-IoT integration approaches and solutions to overcome implementation challenges. The paper also provides condensed high-level knowledge about IoT and Blockchain technologies to identify use cases of Blockchain in IoT systems and networks. 
The remainder of the paper is organized as follows: In Section 2, we briefly explain the Blockchain functionality and describe its advantages and disadvantages. Section 3, presents the integration of Blockchain and IoT systems. In Section 4, we discuss solutions and challenges in this integration. Finally, in Section 5, we provide conclusions and future works.
    
\section{\textbf{Blockchain}}
The Blockchain as derived from its name consists of a chain of blocks. A block is a data structure which allows Blockchain to record the generated and exchanged transactions and each block is linked to the chain by cryptography \cite{wust2018you}. The Blockchain is a distributed ledger which has three fundamental attributes: recorded, transparent, and decentralized. Blockchain forms participants in a P2P distributed ledger to records transactions safely and interact with each other via a trustless method, meaning that there is no need to trust other devices and third parties. All participants keep and update a copy of distributed ledger to check and validate transactions which makes Blockchain transparent and impossible to hack or lost any data \cite{laurence2017blockchain}. Each transaction includes three main components, i.e., the data, the hash, and the hash of the previous block \cite{decuyper2018how}. The data and hash can be defined as follow:

\begin{itemize}
  \item \relax \textbf{\textit{Data:}} The data which is collected inside a block. There can be different data types, depending on the Blockchain applications, for instance, Bitcoin Blockchain stores the transaction information such as the sender, receiver and the number of coins.
  \item \relax \textbf{\textit{Hash:}} The hash is a function that converts a block and all of its contents to a unique fixed-length output which can be interpreted as a fingerprint of the block. Blockchain determines hash once a block created. Modifying the contents of a block will change the hash. Hashes are very useful to detect block tampers. Once the block fingerprint changes, it will be no longer considered as the same block. Hash algorithms take the variable length input string and give out a fixed length output. For instance, Bitcoin uses SHA256 as a hashing algorithm.
\end{itemize}

\begin{figure}[!htbp]
\centering 
{\includegraphics{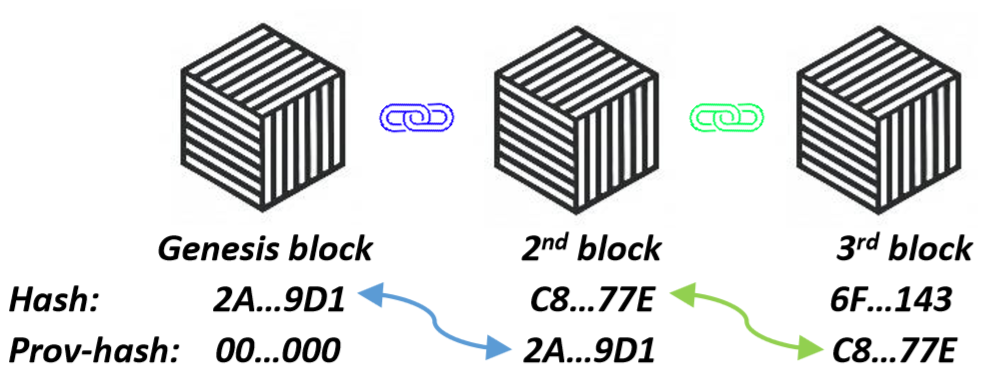}}
\caption{{Blockchain Hash mechanism}}
\label{figure1}
\end{figure}

Each block in the network records the hash of the previous block. This leads to a chain of blocks with enhanced security. For example, in  Figure~\ref{figure1}, there is a chain of three blocks. Block 3 points to block 2 and block 2 points to block 1 using the hashes of previous blocks \footnote{The first block is a bit special because it cannot point to previous blocks. The first block is called the Genesis block.}. If hackers tamper the second block data, the related block hashes changes. This makes the third block and all subsequent blocks invalid because they have not stored a valid hash of the previous block \cite{decuyper2018how}. 

Moreover, any user has two keys: a public key which is known to other users to encrypt their transactions and a private key to read encrypted transactions by the user. Therefore, asymmetric cryptography is used to decrypt the message encrypted by the corresponding public key \cite{ferrag2018blockchain}.

In a P2P network, control and responsibility spread out among lots of different peers, which improve network security. Blockchain utilizes a P2P distributed ledger to eliminate the centralized database risks by storing data across its network and lets everyone to join it. When a node connects to this network, it obtains a full copy of the Blockchain that can later be used to verify if everything is still in order \cite{norman2017blockchain}. A node can be any electronic device, including a computer, phone, a printer or even a fridge, as long as connects to the internet. All nodes have equal importance on a Blockchain. However, each node has different tasks in making a Blockchain. Nodes and their roles can be categorized as follows \cite{norman2017blockchain}:
\begin{itemize}
  \item \relax \textbf{Light node:} Store some of the information recorded on a Blockchain.
  \item \relax \textbf{Full node:} Store a copy of all of the information recorded on a Blockchain. 
  \item \relax \textbf{Mining or forging node:} Process transactions, put them into blocks, add blocks to a Blockchain, approve and broadcast joined block to the network.
\end{itemize}

  These nodes work together to manage, secure and, expand the blockchain. Users in the Blockchain network utilize mining nodes to creating new blocks, verifying their information and adding them to a distributed ledger by executing the consensus algorithm as below \cite{ferrag2018blockchain}: 

\begin{itemize}
  \item \relax User utilizes its private key to sign a transaction and advertises it to its peers. 
  \item \relax User peers validate the received transaction and advertise it over the network.
  \item \relax All the involved participates commonly verify the transaction to meet a consensus agreement. 
  \item \relax Miners add the valid transaction into a time-stamped block and broadcast it again into the network.
  \item \relax After verifying the advertised block and matching its hash with the previous block, this block joins the Blockchain.
\end{itemize}
  
  Consensus protocols are one of the most important and revolutionary aspects of Blockchain technology. Consensus protocol contains rules and verification procedures to validate data which lets the devices around the world, to agree about adding data to the Blockchain \cite{bashir2017mastering}. Based on Blockchain requirements, a large variety of consensus protocols exist. The four main consensus protocols are \cite{bach2018comparative}:

\begin{itemize}
  \item \relax \textbf{\textit{Proof of Work:}} Consists of solving a complex mathematical problem to add a new block to the chain. The process is costly and time-consuming but once solved the solution can be easily verified by other participants. Miners solve a problem, publish the solution and add the new block to the chain that will be spread over the network to be verified by all participants. This process can simultaneously happen in different parts of the network. When peers plan to add a new block, they have to check the branch size and choose the most accumulated work (the longest chain) which is assumed to be the valid one \cite{gupta2018proof}. The proof-of-work defines one CPU one vote approach as a solution for representation problem in majority decision-making. If the majority of CPU powers belong to honest nodes, they control the decision with the fastest growing chain. To modify a past blocks, an attacker would have to redo the proof-of-work of the block and all blocks after it and provides higher CPU power than the honest nodes \cite{nakamoto2008bitcoin}.
  
  \item \relax \textbf{\textit{Proof of Stake:}} Similar to PoW, it attempts to provide consensus. In the PoS the originator of next block is chosen based on the various randomized combination of minors cryptocurrencies resources and the duration that they hold their resources. Contrary to PoW miners that may not have cryptocurrency and only attempt to maximize profits by increasing computational power, PoS miners defend Blockchain network to protect their wealth and profits. As long as the stake is higher than the transaction fees, participants can trust them to do their job correctly \cite{vashchuk2018pros}.
  
  \item \relax \textbf{\textit{Practical Byzantine Fault Tolerance:}} Practical Byzantine Fault Tolerance (PBFT) idea derives from a story about a group of generals, independently commanding a section of the Byzantine army, surrounding a city which they intended to capture. The most important thing is that all generals reach a mutual decision to attack or retreat. The Byzantine problem becomes even more complicated when disloyal generals, votes for an irrelevant strategy \cite{joshi2018survey}. The Byzantine consensus algorithm determines new blocks in rounds and selects the sponsor to advertise an uncorroborated block. The transaction validation includes three steps and in all the phases, the node enters the next stage only after obtaining 2/3 of all network nodes vote \cite{joshi2018survey,castro2002practical}: 
  \\1) Pre-vote step: validators indicate the need to broadcast a block for pre-voting. It is possible to skip this step if the validators believe it is unnecessary for a particular transaction and they can directly approve the pre-voting of a block or transaction by gaining 2/3 votes from the network. 
  \\2) Pre-commit step: If PBFT neglected the pre-vote step; the pre-commit phase goes through the tedious voting phase for broadcast and validation. Once the block receives 2/3 votes for the pre-commit step, it enters the commit phase.
  \\3) Commit step: a node validates a block or transaction and broadcasts a commit for it. This phase accepts the block or transaction validation with 2/3 votes. 
  \item \relax \textbf{\textit{Delegated Proof of Stake:}} Delegated Proof of Stake is one of the fastest, efficient, decentralized, and most flexible consensus models available. DPOS leverages the power of stakeholder approval voting to resolve consensus issues in a fair and democratic way. All network parameters, from fee schedules to block intervals and transaction sizes, can be tuned via elected delegates. The people who hold the particular cryptocurrency will be able to make votes by their token to choose who runs the network. Deterministic selection of block producers allows delegates to confirm transactions in an average of just a second. Perhaps most importantly, the consensus protocol is designed to protect all participants against unwanted regulatory interference \cite{larimer2018delegated}.

\end{itemize}

The smart-contracts are another relevant feature of Blockchain which includes self-executing programs with the terms of the transaction between users. They promise low transaction fees compared to traditional systems that require a trusted third party to enforce and execute the terms of an agreement. They contain a set of code and data that store in a particular Blockchain address and devices can call public functions via this address. Smart contracts give autonomy, trust, backup, safety, saving money, and accuracy to the Blockchain. Even Bitcoin allows some limited set of smart contracts to execute. Ethereum \cite{wood2014ethereum} was the first Blockchain platform which supports arbitrary code execution on the Blockchain \cite{alharby2017blockchain}.
Blockchain can be classified either as private or public which provide a certain level of immunity against faulty or unwanted users for the ledger. The main differences between private and public Blockchains lie in the execution of the consensus protocol, the maintenance of the ledger, and the authorization to join the P2P network. In a private Blockchain, the centralized trusted authority that manages the authentication and authorization process selects the miners. In a public Blockchain, there is no intervention of any third party for the miner selection and joining of a new user to the Blockchain network \cite{fernandez2018review} . Private Blockchains possess some advantages in comparison with public ones such as \cite{buterin2015public}:

\begin{itemize}
	\item \relax Companies can change the rules of a Blockchain, return transactions, and adjust balances.
	\item \relax The known validators protect Blockchain from a majority attack risk.
	\item \relax The cheaper transactions due to less processing power consumption of fewer validator nodes.
	\item \relax Provide a higher level of privacy for reading restricted permissions.
\end{itemize}

\subsection{Blockchain benefits} Considering the information provided in the previous section, the most important benefits of Blockchain technology are \cite{dorri2017Blockchain,wust2018you,laurence2017blockchain}:

\subsubsection{Security} Big companies may put millions of their customers at serious risk if they couldn't provide a secure centralized database. Blockchain uses a distributed ledger to secure its information and protects them against the failure of a centralized decision maker. Furthermore, decentralization guarantees that data remains secure even if one of these devices/nodes fails. Blockchain suggests a high level of security to each individual user as it eliminates using the passwords and online identities by employing powerful cryptography. It provides an address and associated crypto-assets through a combination of public and private keys, making the users identities to not have a direct association with their addresses \cite{joshi2018survey}.

\subsubsection{Transaction Verifiability} Blockchain allows any participant to confirm the integrity of transactions. In a centralized network, the central entity provides the correct state of observers instead of verifying that all state transitions were executed correctly or not. In a distributed ledger, the restricted set of participants serve as verifiers or miners which confirm any changes, while other participants can check the changes as an observer. The ability to validate the transaction by themselves enhances the Blockchain security and reliability \cite{wust2018you}.

\subsubsection{Transparency} Blockchain as a distributed ledger provides data transparency by sharing the same documentations to all participants. These documentations are the immutable data accessible by all Blockchain members which can be only updated by a consensus mechanism. Thus, data transparency on a Blockchain creates a more accurate, consistent database to protect essential data. The participants' level of access to information can change from one to another based on their permission \cite{reyna2018Blockchain}.

\subsubsection{Privacy} There is an inherent trade-off between privacy and transparency. Achieving data privacy in the centralized architecture is easier than the transparent distributed system. However, Blockchain does not need any integrity for the network layer to guarantee the protection of information from unauthorized changes. Cryptography concealed the user identity which makes it arguably impossible to determine the identity accounts owner \cite{wust2018you}.

\subsubsection{Trustless} In Blockchain, participants run consensus protocols to agree on what should be unanimously and securely added to the distributed ledger. Blockchain can verify ownership of anything entirely without the need for a central authority. Smart contracts execute automatically once their terms met. This Blockchain feature eliminates the disputing contracts and contributes to its trustless nature. It is not a case of whether a third party is trusted to carry out tasks, as it is an automated and immutable system in which there is no trust required \cite{atlam2018Blockchain}. 

\subsection{Blockchain challenges:}
Although the fundamental concept of Blockchain is simple, its implementation faces numerous difficulties. This section presents the main challenges in the implementation of Blockchain.

\subsubsection{Storage capacity and scalability} With the continuous growth of transaction amount, the Blockchain size increases and any nodes in the network needs significant storage resources to store data. Although the full copy of Blockchain just saves in the full-nodes, an oversized Blockchain has a negative impact on network functionality. For instance, it will slow down the propagation speed and increases the users synchronization time, leading to Blockchain unwanted forks. Due to the Blockchain size growth, the validation time increases and that needs more computational power to verify the activities over the network. Transaction validation is a fundamental component of consensus protocol which has a direct impact on the Blockchain network scalability \cite{zheng2017overview}.

\subsubsection{Security} Blockchain technology uses numerous techniques to achieve the highest level of security for transactions. Blockchain employs a combination of public and private key to securely encrypt and decrypt data. Blockchain eliminates the 51\% majority attack and fork problems by determining the longest chain as an authentic block \cite{joshi2018survey}. If a Blockchain participant able to manage more than 51\% of the mining power, majority attack happens and in this situation this particular participant is able to control the consensus in the network. The accelerated evolution of mining pools increases the probability of majority attack which could compromise the integrity of Blockchain \cite{eyal2018majority}. The double-spend attack tries to spend the same money more than once. Upon a successful attack, the victim is left with an invalidated payment while having already delivered the service. Bitcoin users protect themselves from double spending fraud by waiting for confirmations when receiving payments on the Blockchain. Multiple variants of the double-spend attack exist. The race attack does only work for fast payment scenarios, e.g., ATMs, cafes or fast food chains. The user sends an unconfirmed transaction directly to the merchant, who accepts it and do not wait for Blockchain confirmation. Meanwhile, they broadcast a conflicting transaction to the network. As the merchant saw their own transaction first, they are under the illusion of getting paid, while the rest of the network predominantly saw the double-spend first and thus it is likely the merchant will in fact not get paid. The second transaction is more likely to be confirmed, and the merchant is cheated. Furthermore, the Finney attack, Denial of Service (DoS), Man in the Middle (MitM) or Sybil can obstruct the network operation \cite{karame2012double}.

\subsubsection{Anonymity and data privacy} Privacy in Blockchain enables the user to perform transactions without leaking its identification information in the network. The Blockchain transparency compromises data privacy even though there is no direct relationship between transactions and individuals. They can reveal the user identity by checking, auditing and tracing each transaction from the system's very first transaction. Therefore, many applications based on public Blockchain technology require a higher level of privacy, specifically in sensitive data use cases. The Blockchain platform accumulates transactions as encrypted data to enhance data privacy. Consequently, the Blockchain compiler is responsible for translating the generic code into cryptographic primitives that supply information anonymity in transactions \cite{reyna2018Blockchain}. Another approach to tackle data privacy is to store sensitive data outside the chain, referred to as the off-chain solution. This kind of solution supports systems that manage large amounts of data since it would be impractical to store them inside the Blockchain. They are particularly suitable for highly sensitive data systems which need tighter access control, such as healthcare applications. Users can utilize the public Blockchain to store anchor data and verify data without relying on authorities, when protected data safely stored off-chain. These off-chain sources must be fault tolerant and should not introduce bottlenecks or single points of failure \cite{lazarovich2015invisible}.

\subsubsection{Smart contracts} In 1994, Nick Szabo proposed the smart-contract concept. It is a self-executable code that runs on the Blockchain to facilitate, perform and enforce the terms of an agreement. Thus, smart contracts guarantee low transaction fees, high-speed, precision, efficiency, and transparency, compared to traditional systems that require a trusted third party to enforce and execute the terms of an agreement. The Blockchain stores smart contracts and allocate a unique address to identify each contract which let any user operate with them only by sending a transaction to this address \cite{szabo1997formalizing}. The benefits of smart-contracts are not obtained at no cost, as they are vulnerable to a series of attacks that creates several new challenges such as hacking, bugs, viruses or communication failures. Bugs in contract coding are highly critical because of the irreversibly and immutable nature of the system. Mechanisms to verify and guarantee the correct operation of smart contracts are necessary to be widely and safety adopted by clients and providers \cite{delmolino2016step}.

\subsubsection{Legal issue} Like any new technology, Blockchain gives rise to some delicate legal challenges. Most of the related laws are becoming obsolete and need to be revised, especially since the emergence of new disruptive technologies such as Blockchain. The development of new laws and standards can ease the certification of security features of devices, and that may help building the most secure and trusted network. Any Blockchain system that holds personal data, for instance, IoT domain applications, will need to comply with applicable data protection laws. The distributed nature of Blockchain is of concern because of cross-border transactions and the ways to execute regulations among different countries with distinct rules. More importantly, creating large data repositories on a Blockchain gives rise to security breaches. Blockchain operators will need to take cybersecurity seriously to avoid potential regulatory action and reputational damage. The need to add more control components over the network has introduced private and consortium Blockchains. These regulations will have an impact on the Blockchain and IoT future and could disrupt the decentralized and free nature of Blockchain by introducing a controlling, centralized participant such as a country \cite{reyna2018Blockchain,governatori2018legal}.

\subsubsection{Consensus} Consensus consist of two functions: First, it allows Blockchain to be updated while ensuring that every block in the chain is valid as well as keeping participants incentivized and second, it prevents any single entity from controlling or crashing the whole Blockchain system. The consensus aim is to create a distributed network without central authorities with participants who do not necessarily need to trust each other \cite{laurence2017blockchain}. An essential disadvantage of primitive consensus protocol is that PoW makes Bitcoin depend on energy consumption. Moreover, miners solve the PoW algorithm to receive the transaction fee and this has led to the situation where people are building larger mining farms. PoW provides more rewards to people with better and more equipment. Even further miners can come together in mining pools to combine their hashing power and distribute the awards across everyone in the pool which makes the Blockchain more centralized as opposed to its decentralized nature and encourage using massive amounts of electricity \cite{gupta2018proof}. PoS is the most popular alternative of PoW consensus approach in Blockchain. It is established on the fact that those users who own more coins, are more interested in the survival and the correct functioning of the system, and, therefore, are the most suitable to carry the responsibility of protecting the system \cite{vashchuk2018pros}. Validators are not chosen completely randomly to verify transactions. A node has to deposit a certain amount of coins into the network as stake that can be considered as security. There is a direct relationship between the validator stake size and its chances to be chosen as forging validator of the next block. PoS consensus protocol might not seem fair because it supports the rich but in reality, it is fairer compared to PoW. In essence, the difference between PoW and PoS are quite significant. PoS does not let everyone mine new blocks and therefore it uses considerably less energy. It is also more decentralized in comparison with PoW that has mining pools\cite{mougayar2016business}.

\section{ Blockchain \& IoT convergence: }
The Internet of Things (IoT) promises to make our lives more convenient by turning each physical object in our surrounding environment into a smart object. IoT exponential extension in recent years creates fundamental challenges in several aspects such as security, privacy, scalability, and maintainability. IoT devices need to operate on effective architecture even in performing simple tasks such as sensing, processing, data collections and communicating. The Blockchain provides many attractive features, such as decentralization, persistency, anonymity, and auditability. These features make Blockchain a promising solution to address some of the paramount challenges in IoT. IoT applications can commonly use Blockchain to access things and store data. Users must be able to access data remotely from any location by using a secure mean and ensure about the privacy of data stored in the network \cite{rayes2016internet,yang2017survey,he2018consensus,lin2017survey}. According to Gartner investigation \cite{walker2016how} in 2025 IoT industries faced five main issues that will be solved by Blockchain technology as below:
 
  \begin{enumerate}
  \item \relax \textbf{\textit{How will industries connect 50 billion devices by 2020?}} Blockchain can store 2\ensuremath{^{160}} addresses which provide IoT devices addressability. More importantly, the Blockchain P2P ledger creates a direct connection between each device to send their information instead of look through a database of billions of records to find that device.
  \item \relax \textbf{\textit{How will industries create controls for vast numbers of decentralized devices?}} Blockchain sends a cryptographically signed message between devices that no hacker can do a man in the middle (MitM) attack or penetrate in it. A user can send control signals from a central location to other decentralized devices.
  \item \relax \textbf{\textit{How will industries enable P2P communication between globally distributed devices?}} Blockchain provides open P2P connectivity for intra-device communication in a natural fashion. Therefore, it becomes very simple to directly send or receive data over the network.
  \item \relax \textbf{\textit{How will industries provide compliance and governance for autonomous systems?}} The Blockchain is an immutable ledger, in other words, the stored data cannot be deleted or edited using which the governance and compliance of autonomous systems become feasible.
  \item \relax \textbf{\textit{How will industries address the security complexities of IoT landscape?}} Bitcoin has proven over ten years that Blockchain powerful protection method could present the strongest communication security in the world for all of IoT devices.
  \end{enumerate}
Thus Blockchain technology presents sufficient advantages for IoT infrastructure that encourages companies to enhance network-based IoT to Blockchain-based one. The Blockchain technology is identified as the main solution for scalability, privacy, and reliability issues in the IoT paradigm. The integration of Blockchain features and protocols in IoT can provide substantial improvements to many IoT applications, for instance \cite{zheng2017overview,dorri2016blockchain,kshetri2017can}:

\begin{itemize}
  \item \relax \textbf{\textit{Decentralization:}} Blockchain offers an effective mechanism to change IoT centralized architecture to a P2P distributed ledger which ensures scalability and robustness using all participants resources and eliminating many-to-one traffic flows. It will decrease latency and solve a single point of failure problem that exist in centralized models (Song et al., 2018). Blockchain prevents the individual authority by using the majority decisions to validate transactions and add them to the distributed ledger \cite{yang2018blockchain}.
  
  \item \relax \textbf{\textit{Immutability:}} Blockchain distributed ledger is immutable, meaning that any data modification must be verified by the majority of the network nodes. Therefore, Blockchain efficiently protects transactions from adjusting or removing. Immutable ledger employment will enhance security and privacy of IoT systems \cite{boudguiga2017towards}.
   
  \item \relax \textbf{\textit{Identity \& access management:}} Blockchain-based identity and access management systems can be leveraged to strengthen IoT security. Public Blockchains let participants identify every single device and their immutable data. They can implement trusted distributed authentication and authorization of devices in IoT applications. Moreover, IoT devices use private Blockchains to store cryptographic hashes of singular device firmware which creates a permanent record of device state and configuration. This record can be used to verify the genuinity of a given device and whether its software and settings have been tampered. Blockchain-based identity and access management systems can provide stronger defense against attacks involving IP spoofing or IP address forgery. The Blockchain resistances against data alteration in the previous blocks make it impossible for devices to connect to network by covering themselves via injecting fake signatures \cite{novo2018blockchain,huh2017managing}.
  
  \item \relax \textbf{\textit{Resiliency:}} Each node stores a copy of the distributed ledger on its memory that contains all transactions have ever made in the Blockchain to combat attacks more efficiently. Keeping such a massive volume of Blockchain data at each IoT node will increase system demand to share information which adds some additional processing, storage, and power consumption to reach more resiliencies in IoT devices.
  
  \item \relax \textbf{\textit{Reliability:}} IoT devices can keep their information immutable and distributed over time via Blockchain technology. Blockchain facilitates sensor data traceability and accountability for tracking billions of connected devices, transactions process, and intra-device coordination. This broad functionality in one technology allows IoT manufacturers to save their resources and budgets. Blockchain full redundancy provides a hundred percent uptime and guaranteeing message delivery \cite{liu2017blockchain}.
  
  \begin{figure*}[b!]
  	\centering
  	\includegraphics[width=1.8\columnwidth=1.8]{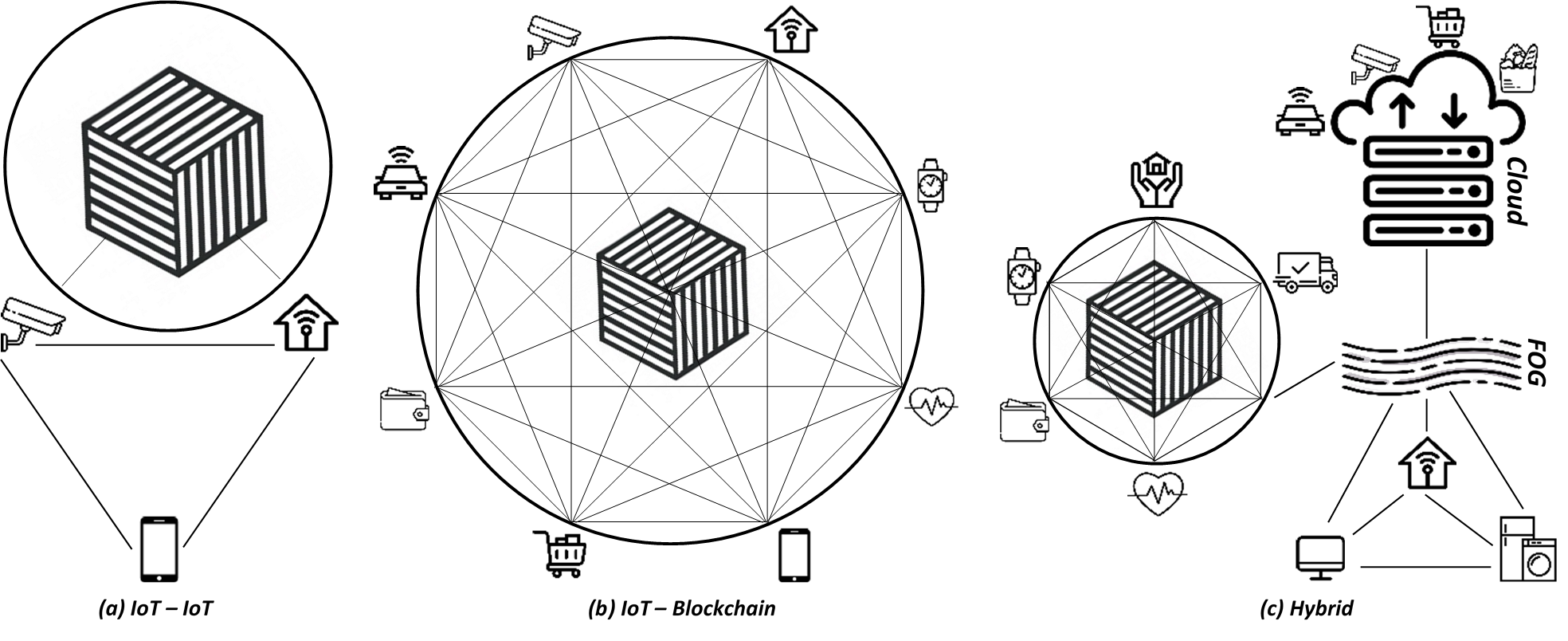}
  	\caption{{ IoT- blockchain convergence approach.}}
  	\label{figure2}
  \end{figure*}

  \item \relax \textbf{\textit{Security:}} IoT devices security flaws typically revolve around three areas: authentication, connection, and transaction. Devices that verify, connect or spend improperly with other devices are all major concerns. IoT system with numerous and heterogeneous devices need Blockchain ability to provide a secure network over untrusted parties. By using Blockchain to manage access to data from IoT devices any attacker would have to bypass an additional layer of security that is underpinned by some of the most robust encryption standards available \cite{halpin2017introduction}. In addition, because there is no centralized authority, single-point failure concerns can no longer be a problem. Therefore, the Blockchain will provide a secure platform for IoT devices by providing a massive amount of trust since the majority of the participants in the network has to reach an agreement to validate transactions. Blockchain can exchange IoT device messages as transactions and validate them by smart contracts. Hack-proof cryptography eliminates attack vectors such as man-in-the-middle (MitM) attacks and all of the other attacks that have been popularized in the last few years when dealing with IoT or industrial internet devices \cite{dorri2017Blockchain}
    
  \item \relax \textit{\textbf{Autonomy:}} Blockchain ability to support IoT devices intra-connection without using any server interposition allow them to communicate autonomously on a worldwide scale. These devices can listen, record or trigger events. The device can transmit a message to other devices based on events that happen autonomously via smart contracts or assets \cite{dorri2016blockchain}. 

  \item \relax \textbf{\textit{Anonymity:}} In Blockchain, both customer and dealer use unknown and unique addresses which privately hold their identities to process the transaction. This feature has been criticized by the government as it increases the use of cryptocurrencies in illegal online markets. However, it could be seen as an advantage if used for other purposes, for example, electoral voting systems \cite{samaniego2016blockchain}.
   
  \item \relax \textbf{\textit{Cost saving:}} Existing Available IoT solutions are expensive because of the high infrastructure and maintenance cost associated with centralized architecture, large server farms, and networking equipment. The total amount of communications that will have to be handled when there are tens of billions of IoT devices will increase those costs substantially \cite{atlam2018Blockchain}.
  
\end{itemize}
  
  Despite increasing agreement on the potential of Blockchain and IoT integration, the main issue about the place where Blockchain would be hosted remains as a disputable topic. Hosting the Blockchain directly on resource-constrained IoT devices are inadvisable due to lack of computational resources, limited bandwidth and their need to preserve power. The cloud and fog are two adapted hosting service platforms for a Blockchain regarding computational resources and latency. While the fog has limited resources and exhibits low latency, cloud-hosted applications can scale out and thus overcome resource constraints at the price of significant latency issues \cite{atlam2017integration}. Based on IoT devices constraint, characteristic, and challenges, the large variety of models proposed for Blockchain and IoT combination in previous researches. These can be classified into three main approaches \cite{reyna2018Blockchain}:

\begin{table*}[!htbp]
	\caption{{ Integration approach state} }
	\label{table1}
	\def\arraystretch{1}
	\ignorespaces 
	\centering 
	\begin{tabulary}{\linewidth}{p{\dimexpr.262\linewidth-2\tabcolsep}p{\dimexpr.1598\linewidth-2\tabcolsep}p{\dimexpr.20\linewidth-2\tabcolsep}p{\dimexpr.1782\linewidth-2\tabcolsep}p{\dimexpr.20\linewidth-2\tabcolsep}}
		\hline 
		& IoT-IoT & Hybrid & IoT-Blockchain & Central Database \\
		\hline 
		Throughput  &
		Low  &
		Medium  &
		High  &
		Very High\\
		Latency &
		Fast &
		Medium &
		Slow &
		Fast\\
		Number of writers  &
		High  &
		High  &
		High  &
		High \\
		Number of untrusted writers  &
		High &
		Low &
		Low &
		0\\
		Data media  &
		BC/IoT devices &
		BC/IoT evices/Fog  &
		blockchain  &
		Cloud\\
		Interaction media  &
		IoT devices &
		BC/IoT devices/Fog &
		Blockchain  &
		Cloud\\
		Consensus Mechanism  &
		Mainly PoW, some PoS &
		PoW, PoS and BFT &
		BFT protocols &
		None \\
		Security  &
		Low &
		Medium &
		High &
		Low\\
		\textit{Consumption Resources} &
		Low &
		Medium &
		High &
		High\\
		\hline 
	\end{tabulary}\par 
\end{table*}

\begin{itemize}
  \item \relax \textbf{\textit{IoT{\textendash}IoT:}} IoT devices usually communicate with each other via discovery and routing mechanisms. Only part of IoT data will be stored in Blockchain whereas the IoT interactions take place without using the Blockchain. This approach is useful in scenarios with reliable IoT interactions with low latency (Figure~\ref{figure2}(a)). 
  \item \relax \textbf{\textit{IoT{\textendash}Blockchain:}} All the interactions and their associated data go through Blockchain, to collect an immutable and traceable record of interactions. This approach is useful in trade and rent scenarios to obtain reliability and security but recording all the interactions increase bandwidth and data resource consumption   (Figure~\ref{figure2}(b)).
  \item \relax \textbf{\textit{Hybrid approach:}} In this approach only part of the interactions take place in the Blockchain and the rest are directly shared between the IoT devices. One of the challenges in this approach is choosing which interactions should go through the Blockchain and providing the way to decide this in run-time. This approach is a perfect way to leverages the benefits of both Blockchain and real-time IoT interactions  (Figure~\ref{figure2}(c)).
\end{itemize}

Researchers (Wust and Gervais, 2017) suggest an algorithm which determines whether Blockchain technology is beneficial or not for any system based on storage demand, writers amount, and trusted third-party requirements. In the cases that companies decided to use blockchain, select a capable integration method depends on requirement appears as another critical point. Table~\ref{table1} presents IoT application requirements such as throughput, data media, latency, security, and resources consumption in different approaches which provides an overall view of their advantage and disadvantages.

The full and mining nodes functionality would be useless in IoT devices due to the restricted power and computation resources. Considering the importance of security in IoT applications, the consensus protocol can be simplified to support more IoT devices. Also, the transaction authentication process can be verified and maintained by lightweight IoT nodes without having to download the entire Blockchain. In any case, Blockchain could be used as an external service to provide secure and reliable storage \cite{reyna2018Blockchain,joshi2018survey}.
    
\section{Blockchain \& IoT integration Challenges and solutions:}
In spite of IoT and Blockchain convergence benefits, this combination is not straightforward. This section studies the main challenges and their related solutions of employing the Blockchain technology which designs for devices with permanent storage and computing resource on the restricted resources IoT devices. The main integration challenges can be summarized as below \cite{reyna2018Blockchain,atlam2018Blockchain}:

\subsection{Blockchain \& IoT Integration Challenges}

\subsubsection{\textbf{Scalability}} The Blockchain size grows with an increasing number of connected devices because of its need to store all transactions to validate them. This is major integration drawback as IoT networks are expected to contain a large number of nodes which can generate massive amount of data in real-time. Additionally, some current Blockchain implementations can only process a few transactions per second. This could be a potential bottleneck for the IoT \cite{zheng2017overview}. To address the Blockchain scalability issue, researchers proposed the Blockchain storage optimization to solve the Blockchain resource challenge via removing old transaction records \cite{bruce2014mini}. Moreover, they worked on redesign Blockchain based on IoT limits. For instance, Bitcoin-NG \cite{eyal2016bitcoin} decouple the common block into the key block for leader election and micro-block to store transactions. Miners are competing to become a leader which responsible for the micro-block generation. 

\subsubsection{\textbf{Security}} The increasing number of attacks on IoT networks and their severe impacts make it necessary to secure IoT devices with Blockchain. This feature maybe creates a severe problem when IoT tools do not operate properly, and corrupted data arrives and remain in the Blockchain. IoT devices should be tested before their integration with Blockchain because of undetectable nature of this problem \cite{roman2013features}. They are likely to be hacked since their constraints limit the firmware updates, preventing them from actuating over possible bugs or security breaches. Moreover, it is difficult to update devices one by one, as required in global IoT deployments. Therefore, run-time upgrading and reconfiguration mechanisms should be placed in the IoT to keep it running over time \cite{reyna2018Blockchain}.

\subsubsection{\textbf{Anonymity and data privacy}} Privacy is a critical concern in IoT. Large amounts of privacy-sensitive data can be generated, processed, and transferred between devices. The Blockchain presents an ideal solution to address identity management in IoT with the ability to hide the identity of the person when sending personal data that protect user data privacy. The problem of data privacy in transparent and public Blockchains has already been discussed, together with some of the existing solutions. The Blockchain transactions use distinct and even dynamic addresses instead of identities. The user anonymity can be revealed by analyzing transactions address which advertised to every participant \cite{he2018consensus}. The IoT devices secured data storage and authorized access, is a challenge since it requires the integration of security cryptographic software to the device taking into account limit resources.
 
\subsubsection{\textbf{Consensus and resource utilization}} Trusted authority in centralized architectures, guarantee the consensus integrity while in the decentralized environment, nodes of the network need to reach consensus by voting, which is a resource-intensive process. IoT devices characterized by relatively low computing capabilities and low power consumption, as well as low-bandwidth wireless connectivity. For instance, Blockchains which utilize PoW as a consensus requires a lot of computing power and consumes a large amount of energy for the mining process. Computationally complex consensus mechanisms are not suitable for IoT scenarios, and the restricted resource should be allocated to reach an agreement. PoS is more likely to be used in IoT, but none of these have yet been deployed in IoT as a standard adoption \cite{atlam2018Blockchain, danzi2018analysis}. A decentralized architecture can reduce the overall cost of the IoT system in comparison to centralized architectures. However, Blockchain as a decentralized architecture suffers from a new type of resource wasting, which poses challenges for its integration with IoT. Resource requirements depend on the particular type of consensus protocol in the Blockchain network. Typically, solutions tend to delegate these tasks to gateways, or any other unconstrained device, capable of providing this functionality. Optionally off-chain solutions, which move information outside the Blockchain to reduce the high latency in the Blockchain, could provide the functionality \cite{reyna2018Blockchain}.

\subsubsection{\textbf{Smart contracts}} Devices can call smart contract functions with addresses or prompt them as application reaction to listening events. They provide a secure and reliable feature for the IoT which record and manage their interactions. Working with smart contracts requires the use of oracles which consist of specific entities that provide real-world data in a trusted manner. Smart contracts executed in individual node whereas simultaneously the code performed by multiple nodes. In other words, instead of using this distribution to execute all tasks, just validation process distributed. Smart contracts should take into account the heterogeneity and limitations which presented in the IoT. Filtering and group mechanisms should be complemented by smart contracts to enable applications to address the IoT problems depending on the context and requirements. Lastly, actuation mechanisms directly from smart contracts would enable faster reactions with the IoT \cite{reyna2018Blockchain}.

\subsubsection{\textbf{Predictability}} Devices in IoT need real-time communication with their environment which means the time used by interactions between things should be predictable and the latency of communication between devices should be bounded. Predictability is even more critical when it comes to healthcare applications based on IoT \cite{bui2011health}.  For example, the transaction finality in Blockchain under many consensus mechanisms, such as PoW and PoS, is probabilistic and the confirmation confidence of the transaction in confusion is also probabilistic. It remains a fundamental challenge to incorporate predictability concerns in the Blockchain architecture \cite{he2018consensus}. 

\subsubsection{\textbf{Legal issues}} The Blockchain connects different people from various countries without having any legal or compliance code to follow which make a serious concern for both manufacturers and service providers. As stated, the lack of regulations for private-key retrieval or reset, or transaction reversion mechanisms creates problems. Some IoT applications envision a global, unique Blockchain for devices but it is unclear if this type of network is intended to be managed by manufacturers or open to users. In any case, Blockchain will require legal regulation. These regulations will have an influence on the future of Blockchain and IoT and maybe disrupt the decentralized and free nature of Blockchain by introducing a controlling, centralized participant such as a country \cite{governatori2018legal}.

\subsection{Blockchain and IoT Integration Solutions}
\begin{figure}[!htbp]
	\centering 
	{\includegraphics{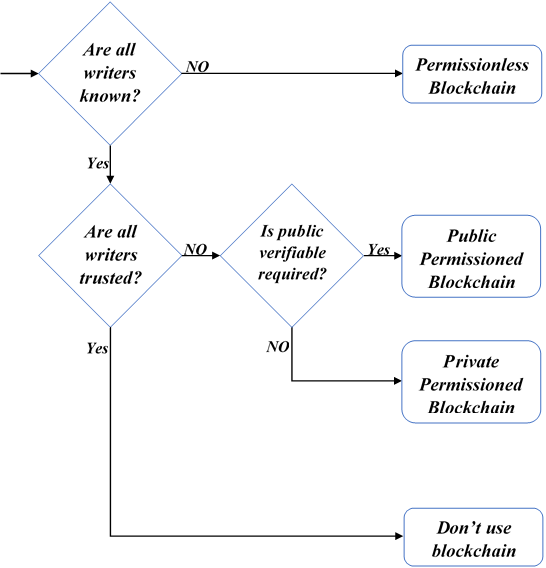}}
	\caption{{Facilitated flowchart of blockchain type selection}}
	\label{figure3}
\end{figure}

The diversity of solutions for Blockchain integration with IoT, and different type of IoT devices and their applications, IoT designers should select an appropriate solution based on their restrictions and requirements. In spite of considerable research on solutions, there has been no comprehensive analysis and resolutions for IoT manufacturers to adopt a suitable Blockchain platform for their integrations. IoT devices need Blockchain to store their state, manage multiple writers, and prevent to hire trusted third party. Figure\ref{figure3} presents a simplified flowchart to determine which kind of Blockchain is suitable for IoT applications \cite{wust2018you,kshetri2017can,song2018Blockchain}.

Table\ref{table2} illustrates Blockchain platforms characteristic and evaluation parameters. This table summarizes important information and evaluation characteristic of well-known Blockchain platforms like Ethereum, Hyperledger fabric \cite{androulaki2018hyperledger}, Multichain \cite{greenspan2015multichain}, Lisk, Neo, and EOS.

\begin{table*}[!htbp]
	\caption{{Blockchain platforms characteristic and evaluation parameters} }
	\label{table2}
	\centering 
	\begin{threeparttable}
		
		\def\arraystretch{1}
		\ignorespaces 
		\centering 
		\begin{tabulary}{\linewidth}{LLLLLLLLLL}
			\hline 
			& Consensus & Block Time & TPS & Scalablity & Security & Privacy & Smart Contract & Pubic/Private & Permission State\\
			\hline 
			Etherum &
			PoW &
			{\texttildeapprox}15s &
			{\texttildeapprox}15 &
			** &
			*** &
			** &
			**** &
			Public & Both\\
			Hyperledger Fabric &
			PBFT &
			{\texttildeapprox}1s &
			{\texttildeapprox}3500 &
			**** &
			*** &
			*** &
			*** &
			Public & Permissioned\\
			Multichain &
			PBFT &
			Adjustable &
			{\texttildeapprox}500 &
			**** &
			**** &
			**** &
			* &
			Private & Permissioned\\
			Lisk &
			DPoS &
			{\texttildeapprox}10s &
			{\texttildeapprox}2.5 &
			*** &
			*** &
			*** &
			** &
			Public & Both\\
			NEO &
			DBFT &
			{\texttildeapprox}15s &
			{\texttildeapprox}1000 &
			*** &
			*** &
			*** &
			*** &
			Public & Both\\
			EOS &
			DPoS &
			{\texttildeapprox}0.5s &
			{\texttildeapprox}10000 &
			*** &
			** &
			**** &
			*** &
			Private & Permissioned\\
			\hline 
		\end{tabulary}\par 
		\begin{tablenotes}\footnotesize 
			
			\item{1- * Low    ** Medium    *** High    ****Very High} 
			\item{2- Both = Permissioned + Permissionless}
		\end{tablenotes}
	\end{threeparttable}
	
\end{table*}

In addition to consensus protocol, block time, and transactions per second (TPS) considered based on their importance for IoT devices to choose an appropriate platform. Any IoT application has perfect knowledge about its restrictions and requirements such as the time-sensitivity, volume of transactions and its resources. This awareness helps IoT devices to define the proper platform. Scalability, security, privacy, and smart contract capability are other metrics that need to be satisfied before platform implementation on IoT devices. Table\ref{table2} introduces these parameters in a qualitative manner to evaluate their performance.Furthermore, building a powerful decentralized network requires some developer tools to work together for smart contracts, faster computation, security, and contract execution to provide a high level of reliability. These protocols are not centralized in data silos and can talk together which enables new use cases to emerge through sharing of data and functionality from multiple protocols in a single application.

\section{\textbf{Conclusion \& potential research direction:}}
With a rapid growth in the number of connected IoT device, many obstacles arise that may slow down the adoption of the IoT across different industries. First, the market for IoT devices and platforms is fragmented, with many standards and many vendors. Second, there are concerns about interoperability, as the solutions implemented often tend to create new data silos. IoT device data often stored in the clouds securely, but they are not protected against compromised integrity devices or tampering at the source. More importantly, the centralized architecture of most IoT solutions require the IoT device owners to trust to these organizations to keep their data safe, to give control over their data and compromise their data if hackers attack the central server.
In contrast, the Blockchain is an emerging technology that can help with IoT systems resiliency. It provides a distributed ledger to avoid centralized architecture challenges and stores data in a secure process via its characteristics. The Blockchain build trust between IoT devices and reducing the risk of tampering with Blockchain cryptography. Moreover, it reduces the cost by eliminating the middlemen and intermediaries overhead. It is intuitive that the Blockchain can provide a promising solution to address many IoT challenges but any convergence between two embedded technologies, create some new issues and obstacles.
IoT devices have limited power and storage resources which cannot handle the resource-intensive distributed ledgers full copy storage, consensus protocol execution and encryption in each node. Moreover, the characteristic of conventional Blockchain should be modified due to IoT requirements such as security, data privacy, the consensus protocol, and smart contracts. One of the main challenges is the heterogeneous solutions that suggest by various types of IoT applications to integrate blockchain with IoT technologies based on their demands and requirements.  In other words, these solutions only focused on specific use cases that may not be suitable to wide range of applications in this area. Therefore, the future research should focus on developing a set of protocols and standards that can support the basic and essential requirements of all IoT applications instead of introducing application-specific IoT networks.



%

\bibliographystyle{IEEEtran}
\bibliography{article}

\end{document}